\documentclass[lettersize,journal]{IEEEtran}
\usepackage{amsmath,amsfonts}
\usepackage{algorithmic}
\usepackage{algorithm}
\usepackage{array}
\usepackage[caption=false,font=normalsize,labelfont=sf,textfont=sf]{subfig}
\usepackage{textcomp}
\usepackage{stfloats}
\usepackage{url}
\usepackage{verbatim}
\usepackage{graphicx}
\usepackage{epstopdf}
\usepackage{cite}
\usepackage{bm}
\usepackage{stfloats}
\usepackage{caption}
\usepackage{subfig} 
\usepackage{array,color}
\usepackage{float}
\usepackage{amssymb}
\usepackage{adjustbox} 
\pdfoutput=1
\allowdisplaybreaks[1]

\begin{document}
	
	\title{Addressing the Mutual Interference in Uplink ISAC Receivers: A Projection Method}
	
	
	
	\author{Zhiyuan Yu, Hong Ren, \IEEEmembership{Member, IEEE}, Cunhua Pan, \IEEEmembership{Senior Member, IEEE}, Gui Zhou, \IEEEmembership{Member, IEEE},
		
		Ruizhe Wang, Mengyu Liu, Jiangzhou Wang, \IEEEmembership{Fellow, IEEE}
		\thanks{The work of Hong Ren was supported in part by the National Natural Science Foundation of China under Grant No. 62471138. The work of Jiangzhou Wang was supported in part by the National Natural Science Foundation of China under Grant No. 62350710796. The work of Cunhua Pan and Hong Ren was supported in part by the Fundamental Research Funds for the Central Universities under Grant 2242022k60001.
			Z. Yu, H. Ren, C. Pan, R. Wang,  M. Liu, J. Wang are with National Mobile Communications Research Laboratory, Southeast University, Nanjing, China. (e-mail:\{zyyu, hren, cpan, rzwang,  mengyuliu, j.z.wang\}@seu.edu.cn). G. Zhou is with the Institute for Digital Communications, Friedrich-Alexander-University Erlangen-N\"{u}rnberg (FAU), 91054 Erlangen, Germany (email: gui.zhou@fau.de). 
			
			\emph{Corresponding authors: Hong Ren and Cunhua Pan.}
		}	
	}
	\maketitle
	\vspace{-1cm}
	\begin{abstract}
		Dual function radar and communication (DFRC) is a promising research direction within integrated sensing and communication (ISAC), improving hardware and spectrum efficiency by merging sensing and communication (S\&C) functionalities into a shared platform. However, the DFRC receiver (DFRC-R) is tasked with both uplink communication signal detection and simultaneously target-related parameter estimation from the echoes, leading to issues with mutual interference.  {\color{black}In this paper, a projection-based scheme is proposed to equivalently transform the joint signal detection and target estimation problem into a joint signal detection process across multiple snapshots.} Compared with conventional successive interference cancellation (SIC) schemes, our proposed approach achieves a higher signal-to-noise ratio (SNR), {\color{black}and a higher ergodic rate when the radar signal is non-negligible.} Nonetheless, it introduces an ill-conditioned signal detection problem, which is addressed using a non-linear detector. By jointly processing an increased number of snapshots, the proposed scheme can achieve high S\&C performance simultaneously.
	\end{abstract}
	
	
	\vspace{-0.3cm}
	\section{Introduction}
	Integrated sensing and communication (ISAC) has received much attention from both academia and industry.  In general,  research in ISAC systems focuses on two main directions: radar communication coexistence (RCC) and dual function radar and communication (DFRC) systems \cite{8288677}. In RCC, communication and radar systems are typically distinct entities, often requiring the exchange of side information, such as sharing the channel state information (CSI) of interference channels. Conversely, DFRC systems are expected to attain a higher level of integration and coordination gains by integrating sensing and communication (S\&C) functionalities into a shared platform.
	
	However, the receivers of the DFRC systems tend to be more complex than those of the RCC system. Specifically, receivers often contend with concurrent S\&C signals in the uplink ISAC systems.  In RCC-receiver (RCC-R),  where radar and communication functions are executed on separate platforms, the receiver design only needs to mitigate the radar interference for communication signal detection \cite{8233171}, or vice versa \cite{9420308}.  In contrast, the DFRC receiver (DFRC-R) is tasked with both uplink communication signal detection and simultaneously target-related parameter estimation from the echoes. These simultaneous occurrences of S\&C tasks result in severe mutual interference, posing challenges in the receiver design.
	
	Several contributions have been made to the uplink DFRC systems\cite{SICOuyang, NOMA_ISAC2,Fan}. The authors of \cite{SICOuyang} demonstrated that despite the mutual interference, the DFRC-R can offer increased degrees of freedom for both S\&C functionalities compared to conventional frequency-division S\&C systems.  Several successive interference cancellation (SIC)-based receiver designs were proposed to mitigate mutual coupling between S\&C signals in \cite{NOMA_ISAC2,SICOuyang}. However, these algorithms may not achieve the desired performance, especially when the S\&C signal possesses comparable strength. The authors of \cite{Fan} proved that the SIC scheme is sub-optimal and an optimal joint signal detection and target estimation scheme was proposed with a tailored minimum mean squared error (MMSE) estimator. However, the proposed algorithm exploits the statistical characteristics of the reflection coefficients, which restricts its applicability to addressing mutual S\&C interference in other scenarios.
	
	Against the above background, our contributions are summarized as follows:
		%
	Firstly, 	we study an uplink DFRC system where the receiver performs uplink communication and sensing service at the same time. A mixed integer least squares (LS) problem is formulated according to the maximum likelihood (ML) estimation of the joint signal detection and target response estimation task. Secondly, we prove that the aforementioned problem can be equivalently transformed into a signal detection problem by the projection operation. Then, several properties of the projection scheme are explored, including the uplink ergodic rate. Finally, by employing the projection scheme, we can achieve high S\&C performance simultaneously at the cost of increased complexity.{

	\vspace{-0.2cm}
	\section{System Model}
	We consider an uplink multiple-input multiple-output (MIMO) DFRC system shown in Fig. 1, which consists of a radar target, an $N_t$ antennas communication user equipment (UE), and a DFRC base station (BS) equipped with $M_t$ transmit antennas and $M_r$ receive antennas.
	
	\begin{center}
		\begin{figure}
			\vspace{-1.cm}
			\centering
			\includegraphics[width=2in]{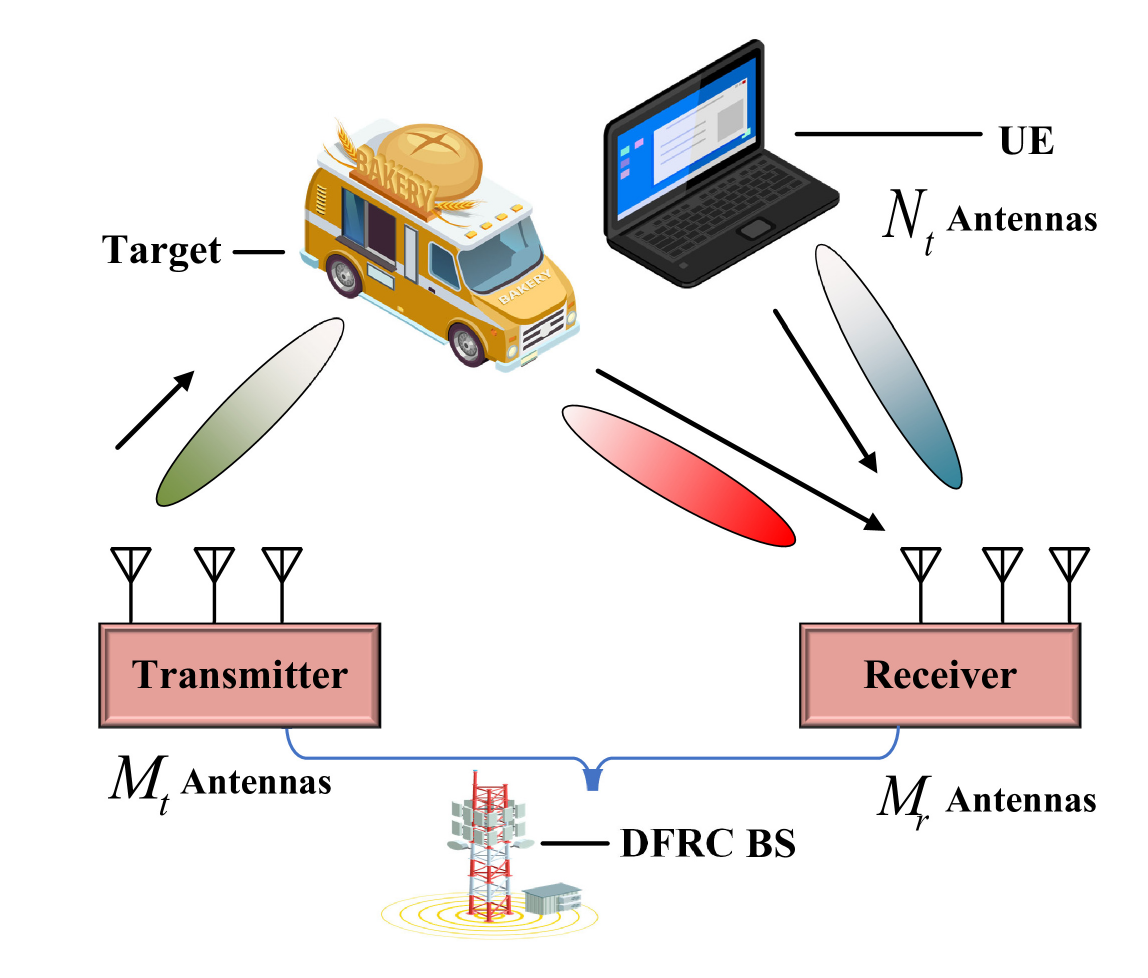}
			\caption{{\color{black}An uplink DFRC system.}}	\vspace{-0.5cm}
			\label{fig1}
		\end{figure}
	\end{center}
	\vspace{-0.9cm}

	Defining $d$ and $\lambda$ as the antenna spacing and the wavelength, the array response vector of a uniform linear array (ULA) with $M$ antennas at angle $\alpha$  is given by $	\mathbf{a}(M, \alpha)=\frac{1}{\sqrt{M}}[1,e^{-j2\pi d \frac{\sin (\alpha)}{\lambda }},{\cdots },e^{-j2\pi d (M-1) \frac{\sin (\alpha)}{\lambda }}]^{\rm{T}}$. 
	
	The received signal at the DFRC BS ${\tilde{\mathbf{y}}}$ is a combination of uplink communication signal and sensing echo. {\color{black}After removing the clutter\cite{OverviewZhang}}, the received signal at the time index $l$, denoted by $ 	{\tilde{\mathbf{y}}}[l]\in \mathbb{C} ^{M_r\times 1}$, is expressed as 
	\begin{align}
		{\tilde{\mathbf{y}}}[l]=&\underbrace{\mathbf{H}_{ {\rm{c}}} \tilde{\mathbf{x}}_{ {\rm{c}}}[l]}_{\textrm{Uplink signal}} +\underbrace{\sum_{p=1}^{P} b_{ p}  
			\mathbf{a}\left(M_{r}, \phi_{p}\right) \mathbf{a}^{\rm{H}}\left(M_t, \theta_{p}\right) {\mathbf{x}}_{ \rm{r}}[l]}_{\textrm{Radar echo}}+\tilde{\mathbf{n}}[l]\\ \nonumber
		\triangleq&\mathbf{H}_{ {\rm{c}}} \tilde{\mathbf{x}}_{ {\rm{c}}}[l]+\mathbf{H}_{ \rm{r}} \mathbf{x}_{ \rm{r}}[l]+\tilde{\mathbf{n}}[l].
	\end{align}
	
	{\color{black}On one hand, the $i$-th entry of the uplink communication signal $\tilde{\mathbf{x}}_{ {\rm{c}}}$ is a modulation symbol taken independently from the discrete constellation $\mathcal{X}$, and the transmit power of the communication signal is denoted by $P_{ \rm{c}}=\mathrm{E}[\|\tilde{\mathbf{x}}_{ {\rm{c}}}\|_2^2]$.}
	In the uplink signal detection task, we assume that the normalized communication channel $\mathbf{H}_{ {\rm{c}}}$ follows $\mathrm{E}[\|{\mathbf{H}}_{ {\rm{c}}}\|_F^2]=N_t M_r$, and the BS decodes the transmit signal $ \tilde{\mathbf{x}}_{ {\rm{c}}}$ according to $\mathbf{H}_{ {\rm{c}}}$ and the received signal. 
	
	{\color{black}On the other hand, the DFRC BS transmits a pre-designed sensing signal and estimates target-related parameters based on the received echo.} { Without loss of generality, we assume that the transmitted signal at the DFRC BS  ${\mathbf{x}}_{  {\rm{r}}}\in \mathbb{C} ^{M_t\times 1}$ satisfies  $ \mathbb{E}\left[{\mathbf{x}}_{  {\rm{r}}}\right]={\mathbf{0}} $ and $\mathbb{E}[\operatorname{Tr}( {{\mathbf{x}}_{  {\rm{r}}}{{\mathbf{x}}^{\rm{H}}_{  {\rm{r}}} }})] =\operatorname{Tr}(\mathbf{R})\le P_{ \rm{r}}$, where $\mathbf{R} \succeq 0$ and $P_{ \rm{r}}$ denote the covariance matrix and the transmit power of the radar signal.} The additive white gaussian noise (AWGN) at the receiver follows the distribution of $\tilde{\mathbf{n}}[l] \sim {\mathcal C}{\mathcal N}\left( {{\mathbf{0}},\sigma^2{{\mathbf{I}}_{M_r}}} \right)$ with the noise power of $\sigma^2$. Here, ${\mathbf{H}}_{ {\rm{r}}}\in \mathbb{C} ^{M_r\times M_t}$ is the target response matrix containing the target-related parameters\{$P, \phi_{p}, \theta_{p}, b_{p}$\}.  Specifically, $P$ denotes the number of the paths, $\phi_{p}$ and $\theta_{p}$ represent the angle of arrival (AoA) and angle of departure (AoD) of the target relative to the receiving and transmitting ULAs, respectively; parameter $b_{p}$ is related to the radar cross-section (RCS) of the target and the distance between the target and the  BS.  These parameters can be acquired if the target response matrix is estimated using an indirect sensing method  \cite[Section III]{OverviewZhang}. 
	
	In the following, we explore the target response estimation and uplink signal detection, taking into account the mutual interference between these tasks. By stacking $L> M_t$ snapshots together, the received signal can be formulated as 
	\begin{equation} \label{vdfz}
		{{\mathbf{Y}}} = {\mathbf{H}}_{ {\rm{r}}}{\mathbf{X}}_{ {\rm{r}}} + {\mathbf{H}}_{ {\rm{c}}}{\mathbf{X}}_{ {\rm{c}}} + {\tilde{\mathbf{N}}},
	\end{equation}
	where ${\mathbf{X}}_{ {\rm{r}}} = \left[ {{\mathbf{x}}_{ {\rm{r}}}[1], \cdots ,{\mathbf{x}}_{ {\rm{r}}}\left[ L \right]} \right] \in \mathbb{C} ^{M_t\times L}$,  ${\mathbf{X}}_{ {\rm{c}}} = \left[ {\tilde{\mathbf{x}}_{ {\rm{c}}}[1], \cdots ,\tilde{\mathbf{x}}_{ {\rm{c}}}\left[ L \right]} \right] \in \mathbb{C} ^{N_t\times L}$, ${\mathbf{Y}} =\left[ {	{\tilde{\mathbf{y}}}[1], \cdots ,	{\tilde{\mathbf{y}}}\left[ L \right]} \right] \in \mathbb{C} ^{M_r\times L}$, and  ${{\tilde{\mathbf{N}}}} =\left[ {{\tilde{\mathbf{n}}}[1], \cdots ,{{\tilde{\mathbf{n}}}}\left[ L \right]} \right] \in \mathbb{C} ^{M_r\times L}$. 
	
	According to the equality $\operatorname{vec}(\mathbf{A} \mathbf{C})=\left(\mathbf{I} \otimes \mathbf{A}\right) \operatorname{vec}(\mathbf{C})=\left(\mathbf{C}^{\mathrm{T}} \otimes \mathbf{I}\right) \operatorname{vec}(\mathbf{A})$, (\ref{vdfz}) can be rewritten as 
	\begin{equation}
		{{\mathbf{y}}} = {\mathbf{A}}_{\rm{r}}{\mathbf{h}}_{\rm{r}} +{\mathbf{A}}_{\rm{c}}{\bf{x}}_{\rm{c}} + {{\mathbf{n}}},
	\end{equation}
	where ${\mathbf{h}}_{\rm{r}}=\operatorname{vec}(\mathbf{H}_{\rm{r}}) \in \mathbb{C} ^{M_r M_t\times 1} $, ${\bf{x}}_{\rm{c}}=\operatorname{vec}(\mathbf{X}_{\rm{c}}) \in \mathbb{C} ^{LN_t\times 1}$,  ${\mathbf{n}}=\operatorname{vec}(\tilde{\mathbf{N}}) \in \mathbb{C} ^{LM_r \times 1} $, ${\mathbf{y}}=\operatorname{vec}(\mathbf{Y}) \in \mathbb{C} ^{LM_r \times 1} $, ${\bf{A}}_{\rm{r}} = {\bf{X}}_{\rm{r}}^{\rm{T}} \otimes {\bf{I}}_{M_r} \in \mathbb{C} ^{L M_r\times M_r N_t}$, and ${\bf{A}}_{\rm{c}} = {\bf{I}}_{L} \otimes {\bf{H}}_{\rm{c}} \in \mathbb{C} ^{L M_r\times L N_t}$. We assume that matrix ${\bf{A}}_{\rm{r}}$ has full column rank, which is typically satisfied when the radar signal matrix ${\bf{X}}_{\rm{r}}^{\rm{T}}$ has full column rank.
	
	Since the AWGN at the receiver follows the distribution of $\mathbf{n}\sim{\mathcal C}{\mathcal N}\left( {{\mathbf{0}},\sigma^2{{\mathbf{I}}_{LM_r}}} \right)$, the probability density function of $\mathbf{y}$ given ${\bf{x}}_{\rm{c}}$ and ${\bf{h}}_{\rm{r}}$ is
	\begin{equation} \label{vzz}
		p({\bf{y}} \mid {\bf{x}}_{\rm{c}}, {\bf{h}}_{\rm{r}})=\frac{1}{(\pi\sigma^2)^{LM_r}} e^{-{\|{{\bf{y}} - {\bf{A}}_{\rm{c}}{\bf{x}}_{\rm{c}} - {\bf{A}}_{\rm{r}}{\bf{h}}_{\rm{r}}}\|_2^{2}  }}.
	\end{equation}
	Therefore, the ML estimation is given by
	\begin{equation} \label{eqq}
		\mathop {\operatorname{argmin} }\limits_{{\bf{h}}_{\rm{r}},\,{{\bf{x}}_{\rm{c}}}\in  \mathcal{X}^{LN_t}}{\left\| {{\bf{y}} - {\bf{A}}_{\rm{c}}{\bf{x}}_{\rm{c}} - {\bf{A}}_{\rm{r}}{\bf{h}}_{\rm{r}}} \right\|_2^2}   .
	\end{equation}
	
	{\color{black}	Solving Problem (\ref{eqq}) involves addressing two primary challenges. First, it is a mixed-integer LS problem,  which minimizes squared errors with both integer and continuous variables. Second,  the equation $\mathbf{AX}+\mathbf{ZB}=\mathbf{C}$ does not have a unique LS solution unless the variables $\mathbf{X}$ and $\mathbf{Z}$ are restricted to a specific region \cite{Liao2005best}. Thus, it is impossible to retrieve ${\bf{X}}_{\rm{c}}$ and $ {\bf{H}}_{\rm{r}}$ in (\ref{vdfz}) without the discrete constellation ${\bf{x}}_{\rm{c}} \in  \mathcal{X}^{LN_t}$.}
	
	\section{Projection-based Solution for DFRC-R}
	\subsection{Conventional SIC Scheme}
	Considering the difficulty of solving Problem (\ref{eqq}), several SIC-based schemes were proposed to address this problem, consisting of two stages \cite{SICOuyang}. In the first stage, the BS initially decodes the communication signal by treating the aggregate interference-plus-noise ${\bf{A}}_{\rm{r}}{\bf{h}}_{\rm{r}}+\mathbf{n}$ as  Gaussian noise. Then, the decoded signal can be obtained by solving the following standard signal detection problem:
	\begin{equation} \label{zbv}
		{\bf{x}}_{\rm{c}}^{\rm{SIC}}=\mathop {\operatorname{argmin} }\limits_{{{\bf{x}}_{\rm{c}} \in  \mathcal{X}^{LN_t}}}{\left\| {{\bf{y}} - {\bf{A}}_{\rm{c}}{\bf{x}}_{\rm{c}} } \right\|^2_2 }  .
	\end{equation}
	In the second stage,  it is assumed that the uplink data has been perfectly decoded and the detected communication signal is subtracted from the superposed signal. The remainder, ${{\bf{y}} - {\bf{A}}_{\rm{c}}{\bf{x}}_{\rm{c}}^{\rm{SIC}} }$, is then utilized to estimate the target response matrix ${\bf{H}}_{\rm{r}}$ by the standard LS method.  
	
	However, the performance of the SIC scheme may be limited due to the following reasons. Firstly, since the radar signal is also treated as Gaussian noise, the communication signal detection problem may have a low signal-to-noise ratio (SNR), leading to a high bit error rate (BER) in the first stage. Secondly, it is challenging to derive the distribution of the residual signal detection error $\bf{x}_{\rm{c}}-{\bf{x}}_{\rm{c}}^{\rm{SIC}}$. Thus, in the second stage, it is difficult to obtain the optimal estimation of the target response matrix. Besides, the residual signal detection error becomes more pronounced due to the high BER in the first stage, significantly impacting the target response estimation performance in the second stage, known as error propagation.{\footnote{{\color{black}Another SIC order is to first perform the sensing tasks by treating the communication signal as noise, which may result in similar performance loss.}}}

	\subsection{Proposed Projection Scheme}
	{\color{black}Inspired by a solution of the mixed integer LS problem proposed in \cite{boyd,Zheng_MIP}, we equivalently transform Problem (\ref{eqq}) into a signal detection problem according to the following theorem:}

	\itshape \textbf{Theorem 1:} \upshape Defining $\mathbf{x}^{\mathrm{ML}}_{\mathrm{c}}$ and $ \mathbf{h}^{\mathrm{ML}}_{\mathrm{r}}$ as the optimal solutions to Problem (\ref{eqq}), solving Problem (\ref{eqq}) is equivalent to solving the following problem:
	\begin{equation} \label{eqvszq}
		\begin{aligned}
			\left(\mathbf{x}^{\mathrm{ML}}_{\mathrm{c}}, \mathbf{h}^{\mathrm{ML}}_{\mathrm{r}}\right)= & \underset{(\mathbf{x}_{\mathrm{c}}, \mathbf{h}_{\mathrm{r}}) \in  \mathcal{X}^{LN_t}\times \mathcal{R}^{M_r M_t}}{\operatorname{argmin}}\left(\mathbf{h}_{\mathrm{r}}-\hat{\mathbf{h}}_{\mathrm{r}}(\mathbf{x}_{\mathrm{c}}) \right)^{\mathrm{H}} \mathbf{\Xi}^{-1} \\
			\times &\left(\mathbf{h}_{\mathrm{r}}-\hat{\mathbf{h}}_{\mathrm{r}}({ \mathbf{x}_{\mathrm{c}}})\right) 
			+\| \mathbf{\Gamma}(\mathbf{y}-\mathbf{A}_{\rm{c}}{\mathbf{x}}_{\mathrm{c}})\|_2^2,
		\end{aligned}
	\end{equation}
	where $\mathbf{\Xi}=\left(\mathbf{A}_{\rm{r}}^{\mathrm{H}} \mathbf{A}_{\rm{r}		}\right)^{-1}$, $\mathbf{\Gamma}=\mathbf{I}_{{LM}_r}-\mathbf{A}_{\rm{r}}(\mathbf{A}_{\rm{r}}^{\rm{H}}\mathbf{A}_{\rm{r}})^{-1}\mathbf{A}_{\rm{r}}^{\rm{H}}$
	and $\hat{\mathbf{h}}_{\rm{r}}$ is a function of ${\mathbf{x}_{\rm{c}}}$, given by 
	\begin{equation} \label{bav}
		\hat{\mathbf{h}}_{\rm{r}}({ \mathbf{x}_{\rm{c}}})=\mathbf{\Xi} \mathbf{A}^{\rm{H}}_{{\rm{r}}}(\mathbf{y}-\mathbf{A}_{\rm{c}}\mathbf{x}_{\rm{c}}).
	\end{equation}

	\itshape \textbf{Proof:} \upshape Please refer to Appendix A. $\hfill\blacksquare$
	
	Since $\mathbf{\Xi} \succeq 0$, the first term in Problem (\ref{eqvszq}) is always non-negative, which implies that its minimum value is zero, and for a given $\mathbf{x}_{\rm{c}}$, its minimizer is ${\mathbf{h}}^{\mathrm{ML}}_{\mathrm{r}}=\hat{\mathbf{h}}_{\mathrm{r}}({ \mathbf{x}_{\mathrm{c}}})$. Notably, the second term in (\ref{eqvszq}) is independent of ${\mathbf{h}}_{\mathrm{r}}$. Thus, minimizing  (\ref{eqvszq}) is equivalent to minimizing the second term in (\ref{eqvszq}) w.r.t. the communication signal $\mathbf{x}_c$, and then deriving the estimated target response according to (\ref{bav}) using the decoded signal.
	
As previously discussed, solving Problem (\ref{eqvszq}) depends on solving Problem (\ref{vbdf}). Several properties of this transformed signal detection problem are then explored:
	\begin{equation} \label{vbdf}
			\mathop {\min }\limits_{{{\mathbf{x}}_{\mathrm{c}}} \in  \mathcal{X}^{LN_t}}  \quad \| \mathbf{\Gamma}(\mathbf{y}-\mathbf{A}_{\rm{c}}{\mathbf{x}}_{\mathrm{c}})\|_2^2
	\end{equation}

	\itshape \textbf{Property 1:}  \upshape Compared to the conventional SIC scheme, the proposed algorithm first projects the S\&C signal into the orthogonal space of $\mathbf{A}_{\rm{r}}$.  Given that $\mathbf{\Gamma}{\bf{A}}_{\rm{r}}=\mathbf{0}_{L M_r\times M_r N_t}$, the interference of the radar signal to the uplink signal detection is completely eliminated. After this projection, we then decode the communication signal and subsequently estimate the target response, following a process similar to the SIC method.
	
	{\color{black}\itshape \textbf{Remark 1:} \upshape  Another major difference between the proposed method and SIC is that the proposed method uses a space-time filter to jointly process $L$ snapshots, whereas in the SIC method, each communication signal is decoded separately.}
	
	\itshape \textbf{Property 2:} \upshape Matrix $\mathbf{G}=\mathbf{\Gamma}\mathbf{A}_{\rm{c}}$ is a singular matrix, and its rank is given by $\operatorname{Rank} (\mathbf{G})=(L-M_t)N_t< L N_t$.

	\itshape \textbf{Proof:} \upshape 	By using the equality $(\mathbf{A B}) \otimes(\mathbf{C D})=(\mathbf{A} \otimes \mathbf{C})(\mathbf{B} \otimes \mathbf{D})$, we have 
	\begin{equation}
		\mathbf{G}\triangleq \mathbf{\Gamma} \mathbf{A}_{\rm{c}}=(\mathbf{I}_L-{\bf{X}}_{\rm{r}}^{\rm{T}}({\bf{X}}_{\rm{r}}^{*}{\bf{X}}_{\rm{r}}^{\rm{T}})^{-1}{\bf{X}}_{\rm{r}}^{*}) \otimes \mathbf{H}_{\rm{c}}.
	\end{equation}
	Defining $\mathbf{P}_{\perp }=\mathbf{I}_L-{\bf{X}}_{\rm{r}}^{\rm{T}}({\bf{X}}_{\rm{r}}^{*}{\bf{X}}_{\rm{r}}^{\rm{T}})^{-1}{\bf{X}}_{\rm{r}}^{*}$ as the orthogonal projection matrix of ${\bf{X}}_{\rm{r}}$, this matrix has $L-M_t$ unit eigenvalues and $M_t$ zero eigenvalues. Applying the rank property of the Kronecker product, we have
	\begin{equation}
		\begin{aligned}
			\operatorname{Rank} (\mathbf{G})&=\operatorname{Rank} (\mathbf{P}_{\perp })\operatorname{Rank} (\mathbf{H}_{\rm{c}})=(L-M_t)N_t. 
		\end{aligned}
	\end{equation}
	Hence, the proof of Property 2 is completed. $\hfill\blacksquare$
	
	
	\itshape \textbf{Remark 2:} \upshape  Defining $\tilde{\mathbf{y}}=\mathbf{\Gamma}\mathbf{y}$, solving the LS problem is equivalent to solving the equation ${\tilde{\mathbf{y}}=\mathbf{G}{\mathbf{x}}_{\mathrm{c}}}$. However, since $\operatorname{Rank}(\mathbf{G})$ is less than the length of ${\mathbf{x}}_{\mathrm{c}}$, the equation has infinite solutions. Therefore, without considering the constellation constraint, we cannot uniquely recover ${\mathbf{x}}_{\mathrm{c}}$ from the observation $\tilde{\mathbf{y}}$, which aligns with the issue described in Problem (5).
	With the constellation constraint, the problem is equivalent to finding the intersection between the solution space of the equation ${\tilde{\mathbf{y}}}=\mathbf{G}{\mathbf{x}}_{\mathrm{c}}$ (independent of $L$)  and the constellation constraint ${{{\mathbf{x}}_{\mathrm{c}}} \in \mathcal{X}^{LN_t}} $. Hence, by increasing the number of snapshots $L$, the intersection becomes narrower, resulting in lower BER.
	
	\itshape \textbf{Property 3:} \upshape The SNR of the transformed signal detection problem in (\ref{vbdf}) is the same as that of the communication-only (comm-only) system, i.e.,
	\begin{equation}
		\textrm{SNR}_{\rm{P}}\triangleq \frac{\mathbb{E}\left[ {\|\mathbf{\Gamma}\mathbf{A}_{\rm{c}}\mathbf{x}_{\rm{c}}\|_2^2}\right] }{ \mathbb{E}\left[ {\|\mathbf{\Gamma}\mathbf{n}\|_2^2}\right]}=\frac{P_{\rm{c}}}{ \sigma^2}=\textrm{SNR}_{\rm{Com}}.
	\end{equation}
	\itshape \textbf{Proof:} \upshape It can be verified that $\mathbb{E}\left[\mathbf{x}_{\rm{c}}\mathbf{x}_{\rm{c}}^{{\rm{H}}}\right]=P_{\rm{c}}/N_t \mathbf{I}_{L N_t}$, $\mathbb{E}\left[\mathbf{X}_{\rm{r}}^{*}\mathbf{X}_{\rm{r}}^{{\rm{T}}}\right]=L \mathbf{R}$, $\mathrm{E}[\|{\mathbf{H}}_{ {\rm{c}}}\|_F^2]=N_t M_r$, and $\mathbb{E}\left[\mathbf{n}\mathbf{n}^{{\rm{H}}}\right]=\sigma^2 \mathbf{I}_{LM_r}$.  Thus, the SNR of Problem (\ref{vbdf}) can be calculated as
	\begin{equation}
		\begin{aligned}
			\textrm{SNR}_{\rm{P}}&\triangleq \frac{\mathbb{E}\left[ {\|\mathbf{\Gamma}\mathbf{A}_{\rm{c}}\mathbf{x}_{\rm{c}}\|_2^2}\right] }{ \mathbb{E}\left[ {\|\mathbf{\Gamma}\mathbf{n}\|_2^2}\right]}=\frac{\mathbb{E}\left[\operatorname{Tr}(\mathbf{\Gamma}\mathbf{A}_{\rm{c}}\mathbb{E}\left[\mathbf{x}_{\rm{c}}\mathbf{x}_{\rm{c}}^{\rm{H}}\right]\mathbf{A}_{\rm{c}}^{\rm{H}}\mathbf{\Gamma}^{\rm{H}})\right]}{\operatorname{Tr}(\mathbf{\Gamma}\mathbb{E}\left[\mathbf{n}\mathbf{n}^{\rm{H}}\right]\mathbf{\Gamma}^{\rm{H}})}\\
			&=\frac{P_{\rm{c}}\mathbb{E}\left[\operatorname{Tr}(\mathbf{GG}^{\rm{H}})\right]}{ \operatorname{Tr}( \mathbf{\Gamma \Gamma}^{\rm{H}})N_t \sigma^2}  =\frac{P_{\rm{c}}\mathbb{E}\left[ \operatorname{Tr}(\mathbf{P}_{\perp } \mathbf{P}_{\perp } ^{\rm{H}}\otimes \mathbf{H}_{\rm{c}}\mathbf{H}_{\rm{c}}^{\rm{H}})\right]}{ \operatorname{Tr}(\mathbf{P}_{\perp } \mathbf{P}_{\perp }^{\rm{H}} \otimes \mathbf{I}_{{M}_r}) N_t \sigma^2} \\
			&=\frac{P_{\rm{c}}\operatorname{Tr}(\mathbf{P}_{\perp } \mathbf{P}_{\perp } ^{\rm{H}}) \mathbb{E}\left[\operatorname{Tr}(\mathbf{H}_{\rm{c}}\mathbf{H}_{\rm{c}}^{\rm{H}})\right]}{\operatorname{Tr}(\mathbf{P}_{\perp } \mathbf{P}_{\perp }^{\rm{H}}) \operatorname{Tr}(\mathbf{I}_{{M}_r})N_t \sigma^2}=\frac{P_{\rm{c}}}{ \sigma^2}.
		\end{aligned}
	\end{equation}
	Hence, the proof of Property 3 is completed. $\hfill\blacksquare$
	
	{\color{black}\itshape \textbf{Remark 3:} \upshape The performance of the comm-only and sensing-only systems can be regarded as the performance upper bound of the uplink DFRC systems. When the signal is projected, the desired signal power and noise power decrease proportionally, thus maintaining the same SNR as the comm-only system. }
	
	As a comparison, the signal-to-interference plus noise ratio (SINR) of the conventional SIC algorithm is given by 
	{
		\begin{equation}
			\begin{aligned}
				\textrm{SINR}_{\rm{SIC}}&\triangleq \frac{\mathbb{E}\left[ {\|\mathbf{A}_{\rm{c}}\mathbf{x}_{\rm{c}}\|_2^2}\right] }{\mathbb{E}\left[ {\|\mathbf{A}_{\rm{r}}\mathbf{h}_{\rm{r}}\|_2^2}\right] + \mathbb{E}\left[ \|{\mathbf{n}\|_2^2}\right]} \\
				&< \frac{\mathbb{E}\left[ {\|\mathbf{A}_{\rm{c}}\mathbf{x}_{\rm{c}}\|_2^2}\right] }{ \mathbb{E}\left[ {\|\mathbf{n}\|_2^2}\right]}=\textrm{SNR}_{\rm{P}}.
			\end{aligned}
		\end{equation}
		

		\subsection{Applying Non-linear Algorithm to Solve Problem (\ref{vbdf})}
In this subsection, we discuss the detailed algorithm to address Problem (\ref{vbdf}), which can be rewritten as
		\begin{equation}\label{optimize for precovzdding4}
			\mathop {\min }\limits_{{{\mathbf{x}}_{\mathrm{c}}} \in  \mathcal{X}^{LN_t}}  \quad  \| \tilde{\mathbf{y}}-\mathbf{G}{\mathbf{x}}_{\mathrm{c}}\|_2^2.			
		\end{equation}
		
		Owing to the rank-deficient property of observation matrix $\mathbf{G}$, solving Problem (\ref{optimize for precovzdding4}) is equivalent to solving a signal detection problem with an ill-conditioned channel matrix, where classical linear decoders, such as zero-forcing (ZF) or MMSE decoders, often yield poor performance. {\color{black} Hence, we adopt a non-linear semi-definite relaxation (SDR) decoder considering the discrete constellation constraint. The implementation details can be found in \cite{SDRLuo,SDRMa}. }
		The overall algorithm of the projection scheme is summarized in Algorithm 1. {\color{black} Note that Step 1 in Algorithm 1 can be conducted offline given the known waveform $\mathbf{A}_{\rm{r}}$, and Step 3 provides similar complexity as the sensing-only system. The primary computational complexity of the proposed method lies in solving Problem (\ref{vbdf}), with a complexity of $\mathcal{O}\left(L^{3.5} N_t^{3.5}\right)$. By contrast, the complexity of the signal detection problem in the comm-only system using SDR is $\mathcal{O}\left(L N_t^{3.5}\right)$. Thus, the proposed projection-based joint signal detection and estimation scheme are generally expected to achieve a better solution while also incurring higher complexity.}
		\begin{algorithm}
			\caption{Projection-based joint signal detection and target estimation} 
			\begin{algorithmic}
				\STATE \textbf{Input}: Received signal $\mathbf{y}$, transmitted sensing signal ${\mathbf{X}}_{{\rm{r}}}$, and perfect CSI of the uplink communication signal ${\mathbf{H}}_{{\rm{c}}}$.
				\STATE \textbf{Output}: Decoded uplink communication signal ${\mathbf{X}}_{\mathrm{c}}$, and estimated target response matrix ${\mathbf{H}}_{\mathrm{r}}$.
			\end{algorithmic}	
			\begin{algorithmic}[1]
				\STATE Project: Calculate the projection matrix $\mathbf{\Gamma}$ and formulate Problem (\ref{optimize for precovzdding4}) accordingly.
				\STATE Decode: Obtain the decoded signal $\hat{{\mathbf{x}}}_{\mathrm{c}}$ by applying the SDR algorithm.
				\STATE Estimate: Estimate the target response matrix with the decoded signal $\hat{{\mathbf{x}}}_{\mathrm{c}}$ according to (\ref{bav}).
			\end{algorithmic}
		\end{algorithm}

		\vspace{-0.7cm}
		\subsection{S\&C Performance Evaluation}
		In the following, we assume that each block contains $L$ snapshots and the Cramér-Rao bound (CRB) of the target response matrix estimation and the ergodic achievable rate {\cite{XiongTIT}} are respectively analyzed in Lemma 2 and Lemma 3.
		
		\itshape \textbf{Lemma 2:}  \upshape In practice, the target response estimation error is acceptable only if most of the communication signal is successfully detected, i.e., block error rate (BLER) $\to 0$. When BLER $=0$, the CRB of the target response matrix estimation is given by  
		\begin{equation}
			\begin{aligned}
				{\mathrm{CRB}}(\mathbf{h}_{\mathrm{r}})=\frac{\sigma^{2} M_{r}}{L} \operatorname{Tr}\left(\mathbf{R}^{-1}\right).
			\end{aligned}
		\end{equation}
		By designing the waveform, the minimum CRB with the constrained transmit power $\operatorname{Tr}(\mathbf{R})\le P_{ \rm{r}}$ is equal to $\frac{\sigma^{2} M_{r}M_{t}^2}{L P_{\rm{r}}}$.
		
		\itshape \textbf{Proof:}  \upshape The minimum CRB can be achieved using the orthogonal waveform, as indicated in {\cite[Section V]{8999605}}.  $\hfill\blacksquare$
		
		{\color{black}	\itshape \textbf{Lemma 3:} \upshape Defining $\lambda_{j}$ as the $j$th eigenvalue of $\mathbf{H}_{\rm{c}}$, the ergodic achievable rate of the comm-only system and uplink DFRC systems under the SIC algorithm are given by 
			\begin{equation}
				C_{{\rm{Com}}}  =\sum_{j=0}^{N_{{t}}-1} \log _{2}(1+{	\textrm{SNR}_{\rm{Com}}} \lambda_{j}\left(\mathbf{H}_{\rm{c}}^{\rm{H}} \mathbf{H}_{\rm{c}}\right) P_{j,1}^{\star}),
			\end{equation}
			\begin{equation} \label{bdf}
				C_{{\rm{SIC}}}  =\sum_{j=0}^{N_{{t}}-1} \log _{2}(1+{	\textrm{SINR}_{\rm{SIC}}} \lambda_{j}\left(\mathbf{H}_{\rm{c}}^{\rm{H}} \mathbf{H}_{\rm{c}}\right) P_{j,2}^{\star}),
			\end{equation}
			where $\{P^{\star}_{j,\{1,2\}}, j=1, \cdots, N_t-1\}$ are the water-filling power solutions of $\lambda_{j}$ in the comm-only systems and the SIC scheme in uplink DFRC systems.
			
			In contrast,   the ergodic achievable rate of the uplink DFRC systems using the projection method is
			\begin{equation}
				C_{{\rm{P}}}  =(1-\frac{M_t}{L})\sum_{j=0}^{N_{{t}}-1} \log _{2}(1+{	\textrm{SNR}_{\rm{P}}} \lambda_{j}\left(\mathbf{H}_{\rm{c}}^{\rm{H}} \mathbf{H}_{\rm{c}}\right) P_{j,1}^{\star}).
			\end{equation}
			
			\itshape \textbf{Proof:} \upshape The ergodic achievable rates of the comm-only system and uplink DFRC systems using the SIC algorithm can be directly obtained according to \cite[Chapter 5]{heath2018foundations}. The ergodic achievable rate of the projection method can be derived by considering the maximum mutual information of the transformed problem:
			\begin{align}
				C_{{\rm{P}}} &\triangleq 	\mathop {\max }\frac{1}{L}	I\left(\tilde{\mathbf{y}}; {\mathbf{x}}_{\mathrm{c}} \mid \mathbf{G}\right)\\ \notag
				&=\sum_{i=0}^{LN_{{t}}-1} \frac{1}{L} \log _{2}(1+{	\textrm{SNR}_{\rm{P}}}\lambda_{i}\left(\mathbf{G}^{\rm{H}} \mathbf{G}\right) P_{i}^{\star}) \\ \notag
				&\overset{(a)}{=}\sum_{i=0}^{L-1}\sum_{j=0}^{N_{{t}}-1} \frac{1}{L} \log _{2}(1+{	\textrm{SNR}_{\rm{P}}}\lambda_{j}\left(\mathbf{H}_{\rm{c}}^{\rm{H}} \mathbf{H}_{\rm{c}}\right)\lambda_{i}\left( \mathbf{P}_{\perp }\right) P_{i,j}^{\star}) \\ \notag
				&\overset{(b)}{=}(1-\frac{M_t}{L})\sum_{j=0}^{N_{{t}}-1} \log _{2}(1+{	\textrm{SNR}_{\rm{P}}} \lambda_{j}\left(\mathbf{H}_{\rm{c}}^{\rm{H}} \mathbf{H}_{\rm{c}}\right) P_{j,1}^{\star}),
			\end{align}
			where equation (a) is due to the fact that the eigenvalue of $\mathbf{A} \otimes \mathbf{B}$ is the product of that of $\mathbf{A}$ and $\mathbf{B}$, and equation (b) is obtained by using the fact that  $\mathbf{P}_{\perp }$ has $L-M_t$ unit eigenvalues and $M_t$ zero eigenvalues. $\hfill\blacksquare$
			
			\itshape \textbf{Remark 4:} \upshape 
			Combining Lemma 2 with Lemma 3, we have the following observations. The SIC scheme tends to achieve the desired achievable rate when the power of the radar signal $P_{\rm{r}}$ is negligible, which also leads to low target response matrix estimation precision. In contrast, our projection scheme can achieve better S\&C performance simultaneously. More interestingly, we observe that increasing the number of snapshots improves both the sensing performance and the ergodic rate. When $L\rightarrow \infty$, the ergodic rate becomes the same as that of the comm-only system, indicating that the impact of the interference can be perfectly canceled.
		}

		\section{simulation results}
		Simulation results are provided to evaluate the performance of the uplink DFRC systems. The DFRC BS is equipped with $M_t=4$ transmit antennas and UE is equipped with $N_t=8$ transmit antennas. {\color{black}Without loss of generality, we assume QPSK modulation for uplink communication.}  We also assume that the power of the communication signal is $P_{\rm{c}}=1$ W, and the noise power is equal to $\sigma^2=-20$ dB. 
		
		\begin{figure}[htbp]
			\centering
			\vspace{-1.2cm}
			\begin{minipage}[t]{0.49\textwidth}
				\centering
				\includegraphics[width=7cm]{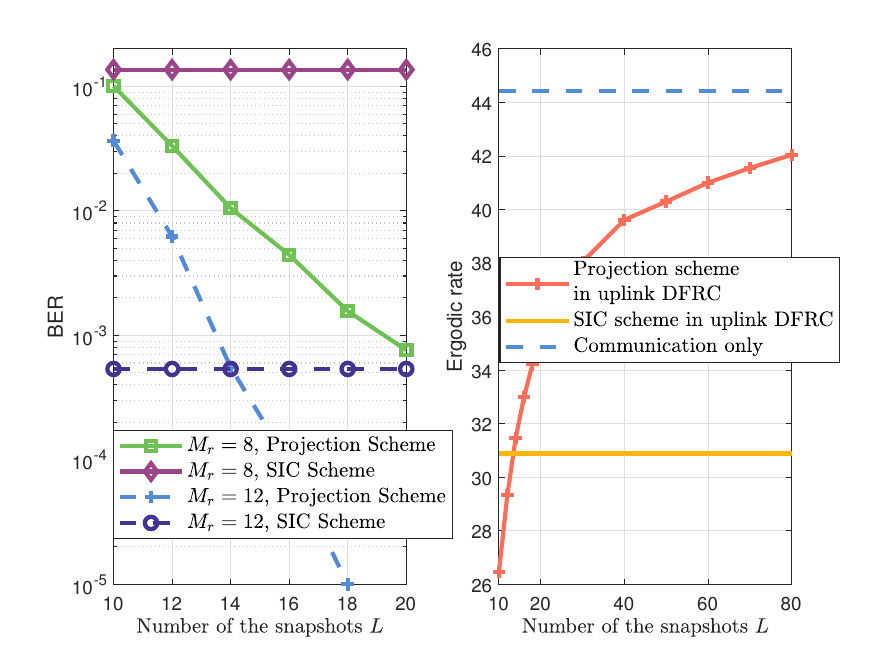}
				\caption{\small{\color{black}BER and ergodic rate versus the number of the snapshots.}}
			\end{minipage}
			\vspace{-0.2cm}
			\begin{minipage}[t]{0.49\textwidth}
				\centering
				\includegraphics[width=7cm]{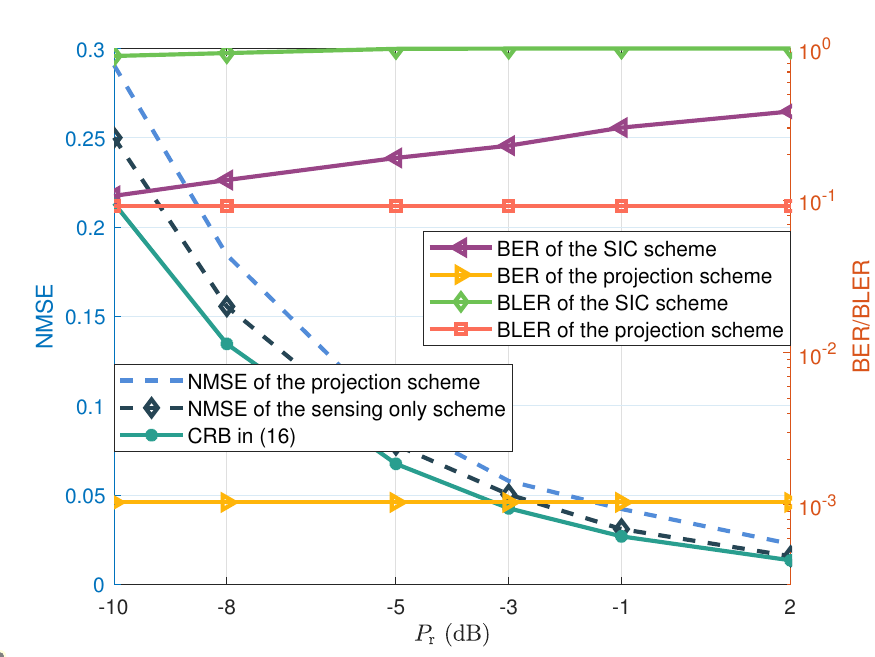}
				\caption{\small{\color{black}S\&C performance evaluation under different $P_{\rm{r}}$}.}
			\end{minipage}
		\end{figure}

		
		In Fig. 2, we set the power of the radar signal to $P_{\rm{r}}=-8$ dB and examine the BER and ergodic rate under varying numbers of the processing snapshots $L$. It is observed that the performance of the SIC scheme is independent of the number of snapshots $L$, but exhibits a higher BER and lower capacity. In contrast, the projection scheme tends to achieve lower BER with an increasing number of $L$, as discussed in Remark 2.  Besides, compared to $M_r=8$, $M_r=12$ exhibits a lower BER due to the increased transceiver antenna ratio. Finally, the ergodic rate of the projection scheme also increases w.r.t. $L$, and approaches that of the comm-only systems if $L$ is large. 
		
		In Fig. 3, we set $M_r=8$ and $L=20$, respectively. The BER, BLER, and the normalized mean square error (NMSE) of the target response matrix are used to evaluate the S\&C performance. The SIC scheme demonstrates worse communication performance, particularly in terms of BLER as $P_{\rm{r}}$ increases. Furthermore, when part of the signal is not successfully decoded, the decoding error significantly influences the sensing NMSE if the communication signal is non-negligible to the sensing signal, resulting in $\textrm{NMSE}>1$ when $P_{\rm{r}}\in [-10,2]$ dB.  In contrast, the BLER and BER of the projection scheme are independent of the power of the sensing signal due to the projection process. With guaranteed communication performance, increasing the sensing power leads to a lower NMSE, approaching that of the sensing-only scheme and the derived CRB. 
	
		\section{Conclusion}
		
		In this paper, we investigated the receiver design of an uplink DFRC system, where S\&C tasks are mutually interfered. We introduced a projection-based approach to effectively handle the ML joint signal detection and target estimation problem. Various characteristics of this method were examined, revealing its ability to simultaneously attain high S\&C performance at the cost of increased complexity.

		\numberwithin{equation}{section}
		\begin{appendices}	
			\section {Proof for Theorem 1}
			We first introduce the following Lemma to prove Theorem 1.
			
			\itshape \textbf{Lemma 1:} \upshape If matrix $\mathbf{A}$ has full column rank, the square error can be equivalently written as
			\begin{equation}
				{\left\| {{\bf{b}} - {\bf{A}}{\bf{x}}} \right\|_2^2}=\|\mathbf{A}(\mathbf{x}-\tilde{\mathbf{x}})\|_2^2+\underbrace{{\mathbf{b}^{\rm{H}}(\mathbf{I}-\mathbf{A}(\mathbf{A}^{\rm{H}}\mathbf{A})^{-1}\mathbf{A}^{\rm{H}})\mathbf{b}}}_{\textrm{Minimal Estimation Error}},
			\end{equation}
			where $\tilde{\mathbf{x}}=(\mathbf{A}^{\rm{H}}\mathbf{A})^{-1}\mathbf{A}^{\rm{H}}\mathbf{b}.$
			
			\itshape \textbf{Proof:} \upshape This lemma can be readily proved by expanding the equation from both sides.
			$\hfill\blacksquare$
			
			Since matrix $\mathbf{A}_{\rm{c}}$ has full column rank, according to Lemma 1, Problem (\ref{eqq}) can be formulated as
			{\small
				\begin{equation}
					\begin{aligned}		
						\underset{(\mathbf{x}_{\mathrm{c}}, \mathbf{h}_{\mathrm{r}}) \in \mathbf{R}^{p} \times \mathbf{Z}^{q}}{\operatorname{argmin}}&\left(\mathbf{h}_{\mathrm{r}}-\hat{\mathbf{h}}_{\mathrm{r}}{ (\mathbf{x}_{\mathrm{c}})}\right)^{\mathrm{H}} \mathbf{\Xi}^{-1}\left(\mathbf{h}_{\mathrm{r}}-\hat{\mathbf{h}}_{\mathrm{r}}{(\mathbf{x}_{\mathrm{c}})}\right) \\
						+&(\mathbf{y}-\mathbf{A}_{\rm{c}}{\mathbf{x}}_{\mathrm{c}})^{\rm{H}}(\mathbf{I}-\mathbf{A}_{\rm{r}}(\mathbf{A}_{\rm{r}}^{\rm{H}}\mathbf{A}_{\rm{r}}))^{-1}\mathbf{A}_{\rm{r}}^{\rm{H}})(\mathbf{y}-\mathbf{A}_{\rm{c}}{\mathbf{x}}_{\mathrm{c}}) \\
						=&\left(\mathbf{h}_{\mathrm{r}}-\hat{\mathbf{h}}_{\mathrm{r}}{(\mathbf{x}_{\mathrm{c}})}\right)^{\mathrm{H}} \mathbf{\Xi}^{-1}\left(\mathbf{h}_{\mathrm{r}}-\hat{\mathbf{h}}_{\mathrm{r}}{(\mathbf{x}_{\mathrm{c}})}\right) \\
						+ &\| \mathbf{\Gamma}^{\frac{1}{2}}(\mathbf{y}-\mathbf{A}_{\rm{c}}{\mathbf{x}}_{\mathrm{c}})\|_2^2,
					\end{aligned}
			\end{equation} }
			where $\mathbf{\Gamma}=\mathbf{I}_{{LM}_r}-\mathbf{A}_{\rm{r}}(\mathbf{A}_{\rm{r}}^{\rm{H}}\mathbf{A}_{\rm{r}})^{-1}\mathbf{A}_{\rm{r}}^{\rm{H}}$ is the orthogonal projection matrix of $\mathbf{A}_{\rm{r}}$. By using the property of the orthogonal projection matrix, we have $\mathbf{\Gamma}^{{1}/{2}}=\mathbf{\Gamma}$.
			
			Hence, the proof of  Theorem 1 is completed.
			\end{appendices} 
			
			\bibliographystyle{IEEEtran}
			\bibliography{IEEEabrv,AMI}

		\end{document}